# The Importance of the Ultra-alkaline Volcanic Nature of the Raw Materials to the Ductility of Roman Marine Concrete

Jackson MacFarlane[1], Tiziana Vanorio[1] and Paulo J.M. Monteiro[2]

[1]Stanford Rock Physics Laboratory, Stanford University

[2]Department of Civil Engineering, UC Berkeley

**Correspondence to:** Jackson MacFarlane, email: jmacfarl@stanford.edu

**Abstract**

Roman-era concrete is the iconic embodiment of long-term physicochemical resilience. We investigated the basis of this behavior across scales of observations by coupling time-lapse (4-D) tomographic imaging of macroscopic mechanical stressing with structural microscopy and chemical spectroscopy on Roman marine concrete (RMC) from ancient harbors in Italy and Israel. Stress-strain measurements revealed that RMC creeps and exhibits a ductile deformation mode. The permeability of specimens from Italy were found to be low due to increased matrix-aggregate bonding. Structural and chemical imaging shows the presence of well-developed sulfur-rich, fibrous minerals that are intertwined and embedded in a crossbred matrix having the chemical traits of both a calcium-aluminum-silicate-hydrate and a geopolymer. This latter likely reflects the ultra-alkaline volcanic nature of the primary source materials. We hypothesize that the fine interweave of sulfur-rich fibers within this crossbred matrix enhances aggregate bonding, which altogether contributes to the durability of RMC.

1. **Introduction**

Among Earth-derived ancient materials, Roman concrete has stood the test of time becoming the embodiment of long-term durability and physicochemical resilience. Such longevity has spurred considerable research across disciplines with interest in understanding the properties responsible for the exceptional stability of Roman concrete over time [1–6]. Physicochemical resilience is inherently connected to how the time-dependent behavior of the macroscopic system is related to the behavior of its microscopic constituents. In the specific case of Roman marine concrete (RMC), the microscopic physical and chemical structures that provide the macroscopic durability remain undiscovered. With modern Portland cement being responsible for 8-9% of global anthropogenic $CO_2$ emissions, there is great demand for long lasting cements that encapsulate environmental goals and energy-efficiency [7]. Given both the long-term durability and the historical significance of RMC [4], understanding the connection between the microstructural and macroscopic properties of RMC is of great interest for both human understanding and technological developments.

Our primary understanding of the engineering of Roman concrete comes from the handbook *De Architectura* by Vitruvius [8], a Roman engineer, who detailed the different components and procedures for the making of marine and terrestrial concretes. Based on the recipe passed on to us, RMC was made from the volcanic ash of Putéoli (i.e., the Roman Pozzuoli in the Campi Flegrei caldera) now known as *pozzolana*, slaked lime and water, to which various types of rubble aggregate were added. Of these ingredients, the sources and composition of the *pozzolana* and aggregates are well constrained; The *pozzolana* is a low calcium (~2%) volcanic ash from ultra-alkaline (potassic) magmatism (alkali > 10%) whereas the composition of aggregates varies, depending on the specific location, being made of either the Neapolotan Yellow Tuff (NYT, a lithified and zeolitized *pozzolana*) or locally quarried rocks [2,4,9,10]. Contrarily, the composition of the lime [4], including the rock used to produce it and water, which could have been ground or seawater, and how the resulting chemical and microstructural fabric contribute to mechanical resilience remain unexplored.

The starting point of this study is the findings from the Roman Maritime Concrete Survey (ROMACONS), being to date the most comprehensive study of RMC [4]. The ROMACONS collected core samples from 11 locations throughout the Mediterranean, ranging from the first century BC to the second century AD, with the objective of providing information on engineering properties, processes, and execution. An in-depth chemical investigation of the solid phases revealed that all of the samples were comprised of a poorly crystalline calcium alumino silicate hydrate (C-A-S-H) matrix, with sulfate, alumina and chloride containing crystalline structures associated with relict lime clasts [2,11]. Within the literature, experimental characterization of the stress-strain response and fracture behavior of RMC refer either to mortar reproductions [4,12] or reporting the ultimate unconfined strength without specifying failure mode or creep behavior [4].

Recent research has shown the significant impact that reaction chemistry has on a materials resistance to fracture and crack propagation as much for concrete [13,14] as Earth materials [15,16]. In fact, not only have pozzolans from calc-alkaline volcanic regions been used as cementitious binders over the centuries [4,17,18] but many cemented ash formations in the subsurface exhibit absorbance of mechanical strain energy (Table 1). The absorbance manifests either as low seismic efficiency in response to large deformation in volcanic areas [16,19] or inefficient fracture height growth during hydraulic fracturing of shale reservoirs [20]. The alkalinity of these natural pozzolans and the durability of the both natural and artificial ash based cements corroborates the controls of reaction chemistry in concrete.

Within engineered cementitious materials, the pathway to increased material durability can be found in ultra-ductile, low-permeability materials that resist high stresses and chemical attack [21–23]. As the engineering of cement chemistry has been refined over time, binder additives and modifiers have emerged as important regulators of the chemistry of ash cementation, which improve concrete performance. Among these additives, alkali-activators have been researched and used in ash-based cements. The presence of alkali facilitates the polymerization of C-A-S-H [24–27], which yields reaction products with an increased resistance to acids and sulfate [28,29], temperature [30,31], and stress [29]. Despite alkali-rich structures being identified in RMC, the link between polymerization of the cement microstructure and durability has not been investigated.

In this study, we investigate the links between microstructural composition and macroscopic durability in RMC. Specifically, we used high-resolution surface and internal imaging, combined with chemical mapping, to characterize the microstructure. Since mechanical and chemical resilience depend on toughness and reactive transport processes, we characterized the macroscopic durability through stress-strain measurements and the transport properties (porosity and permeability). This represents the first comprehensive investigation into the durability of RMC that has combined these measurement techniques in a continuum approach across multi-scales of observations, from the micro to macro scale.

## 2. Materials and Methods

The samples used in this study were collected by the Roman Maritime Concrete Survey (ROMACONS) from the harbors of Cosa, Traiani and Baia (Italy) and Caesarea (Israel), with the sampling procedures described in [4]. The 88 mm diameter cores have visible large aggregates and a heterogeneous matrix of smaller aggregates, white clasts and cement. The initial larger cores were sent to Core Laboratories to be sub-cored into 25.4 mm diameter core-plugs as required for all the physical testing apparatuses. The coring locations for the 25.4 mm diameter core-plugs where chosen to have primarily either cement matrix or aggregate (henceforth referred to as matrix-dominated and aggregate-dominated) so as to assess properties separately. The cored samples were cut approximately 25.4 mm in length using a precision saw and trimmed carefully so as to remove large aggregates near the end surfaces, which would affect the bulk properties of interest.

The stress-strain measurements were performed in a triaxial vessel (Core Laboratory HCH-1.0-AC), which is a biaxial type core holder ($\sigma_1 > \sigma_2 = \sigma_3$) that can be configured for both hydrostatic and triaxial stresses, allowing for independent axial and radial loading. The sample was jacketed with rubber tubing to isolate it from the confining pressure medium (Chevron Rando HD ISO 32). The radial and axial pressures were controlled using a dual piston Quizix QX6000SS. The sample was measured dried with the pore fluid lines open to perform a drained experiment maintaining a pore fluid pressure of zero. The sample was hydrostatically loaded (radial pressure = axial pressure) incrementally from 0 to 5 MPa twice before the radial pressure was held constant at 3 MPa and the axial pressure was increased in a stress-stepping experiment. Stress-stepping was used to measure the creeping strain rate as a function of differential stress as devised by Gasc-Barnier et al. [32]. The experiments were terminated when a sudden loss in pressure and decrease in length was measured. The change in length is measured independently by three linear potentiometers that are connected to the steel end cap providing axial pressure. The voltage change in each potentiometer is proportional to the displacement and the average of the three measurements is used to calculate the axial length change. The strain rate was calculated as a linear fitting of the strain as a function of time at each constant stress step.

Porosity measurements were taken using a helium porosimeter. The porosimeter uses the gas expansion technique based on Boyle's and Charles' law for the expansion of gases [33]. This allows for the measurement of the grain and pore volume, and hence for the calculation of porosity. All measurements of permeability are made on a CoreTest Inc. Automated Permeameter 608 using the pulse-decay technique [34]. This technique is particularly suitable for low permeability specimens and consists of producing an unsteady state decay curve by confining the sample at a pressure of 500 kPa, applying a helium pressure pulse upstream, and then allowing the pressure to decay across the sample to atmosphere. That pressure decay curve is continuously recorded with time and automatically fitted to provide the Klinkenberg corrected permeability. The experimental error in both the porosity and permeability measurements are approximately 1%.

Micro-CT scans were performed on a Zeiss Xradia Versa 520 at the Stanford Nano Shared Facilities. Whole plug scans were obtained using a polychromatic beam centered at either 80 or 140 keV, the lowest energies of which were removed by Zeiss filters LE5, LE6 or HE1. Due to variable X-ray absorption between samples, the combination of beam energy and filter was chosen for each sample to achieve transmission values between 20-35%. The projections were optically magnified 0.4x prior to detection. Each scan contained a portion of the sample within the field of view and was independently reconstructed using the supplied Zeiss software, with which the center shift and beam hardening constraints were tuned. For samples with heights greater than the field of view, tomographic datasets were created by vertically stitching successive reconstructions. Voxel resolutions ranged between 27 to 35 µm. Reconstructions were explored and analyzed using Dragonfly 3.0 and ImageJ [35,36]. After failure, sample Cosa2 was removed from the pressure vessel while still contained within its rubber jack to allow for a second micro-CT scan. The micro-CT shown in Figure 7 was performed by Zeiss using a Zeiss Xradia Versa 520. For this image the beam was centered at 150 keV and the voxel size is 75 µm.

SEM images were collected using two systems. The first system was using a Hitachi S-34N VP that allows for imaging without carbon coating. This allows for imaging in between measurements that require surface analysis. Imaging was performed at an acceleration voltage of 15 keV and vacuum pressure of 50 Pa. High-resolution surface images of the samples with elemental analysis were acquired using a JEOL JXA-8230 "SuperProbe" electron microprobe. Sample slices were carbon coated to reduce surface charging. Backscatter election (BSE) imaging was performed at an acceleration voltage of 15 keV and probe current 20 nA. Elemental analysis was performed using energy-dispersive spectroscopy (EDS) imaging with a silicon drift detector. Point analysis was performed with a 15 second dwell time.

The micro-XRF analysis was performed by SIGRAY using a Molybdenum source running at 20W with Moly Optics. The source was 90 degrees to the sample surface and scanning was done with a 10µm step size and 0.1 seconds dwell time/pixel. FTIR-ATR analysis was performed using a Nicolet iS50. Samples were prepared by taking small amounts from cores from Caesarea and Cosa and crushing them into a fine powder. Additional samples from Cosa were taken by selectively sampling from an area including a large relict lime clast and without a large clast. Powdered samples were analyzed using a KBr beam splitter with 32 scans taken per sample and a single background taken before each sample. The spectra were analyzed using Spectragryph v1.2.10. A constant baseline was taken to align spectra and then they were normalized to the largest peak.

In addition to samples of Roman concrete, we created RMC analogous mortar samples in the laboratory by mixing pozzolana, slaked lime, and water, mimicking the process described by Vitruvius for creating Roman cement. Sample mixtures were prepared under different pressure conditions to obtain a compaction trend spanning a range of both porosity and permeability (Table S1). The pozzolana was collected from a quarry within the Campi Flegrei caldera, prior to use it was humidified and sieved to 1.7 mm. The cement slurry was poured into a cylindrical mold and compressed with varying force varying from a 5 lb. mass to a 2 MPa uniaxial press and time varying from 3 to 24 hours. After compression the samples were removed from the split cube and placed in a steam chamber heated to 80°C for 14 or 28 days, creating a high relative humidity for preventing de-hydration and hence, hydrate mineral formation. After curing the samples were removed from the steam chamber and dried in an oven at 80°C for 5 days to halt the hydration reaction.

## 3. Results

When placed under triaxial compression, the mortar-dominated samples from Cosa exhibited ductile deformation modes, with large axial strain relative to the stress (Figure 1a). The samples exhibited rapid stress drops at deviatoric stresses between 2.3-2.9 MPa (radial pressure held constant at 3 MPa), experiencing axial strains of more than 9%. During the stress-stepping experiments the samples continued to deform axially during the constant stress phases due to short-term creep (Figure S1). The creeping strain rate was found to increase by three orders of magnitude as the deviatoric stress was increased from 0-3 MPa. Despite the heterogeneity of the samples, and slightly different porosities (46.3% and 51.4%), the samples show similar ductile deformation and creeping strain rates. Time-lapse imaging of the microstructure using X-ray micro-Computed Tomography (micro-CT) prior to and after the stress-strain experiment (Figure 2a-b) provides the opportunity to directly observe a process to reveal how it mechanistically unfolds, thus detailing information on how the sample has deformed. Imaging reveals that the strain in the sample is accommodated by pore collapse (matrix densification), matrix rearrangement (grain rotation and relative movement) and isolated micro-fracturing. Micro-fractures within the samples (Figure 2c) are concentrated around the outer edges of the sample, inside relict lime clasts and where aggregate debonding occurs. Micro-fractures remain isolated and do not merge into through-going cracks.

Scanning Electron Microscopy (SEM) imaging shows the matrix of the cores from both Baia and Cosa to be pervasively intruded by fibrous features (Figure 3), which are well developed and range in length from 1-1000 μm. High magnifications of the microstructure show the fibers to be finely embedded and intertwined in the matrix of the mortar (Figure 3b) and sprouting from the matrix throughout the pore spaces like threads interlacing each other in a finely woven cloth (Figure 4a). The spatial distribution of the pore-intruding fibers can be seen in the micro-CT scans, which reveal that the fibers are within moldic porosity (pore space that is left behind by the in-situ reaction of lime) throughout the mortar. The moldic pores are bordered by reaction rims (Figure 4b), which remain as evidence of the reaction. In some instances, pores still contain relict lime clasts that appear structurally heterogeneous and are surrounded by incipient reaction halos (Figure 4b and S2). These fibrous features were not identified in cores from Caesarea neither through SEM nor X-ray imaging.

The Electron Probe Micro Analysis (EPMA) mapping identified the fibers to be rich in sulfur, calcium, and aluminum. Conversely, the matrix shows a richness in calcium, aluminum and silica (Figure 5b, c and S3). This chemistry correlates, respectively, with the identification of ettringite, a calcium-sulfoaluminate-hydrate [4], and Al-tobermorite, a crystalline analogue of C-A-S-H, in the matrix of RMC [2,11]. Furthermore, both the matrix and fibers contain alkalis such as sodium and potassium (Figure 5d-f, Table S2) with concentrations in the matrix being heterogeneously distributed and negatively correlated with that of calcium (Figure S4). To identify molecular components and structures of the matrix, we performed Fourier-transform infrared spectroscopy (FTIR) of RMC. The spectrum shows a main Si-O bonding band between 950 and 1050 cm$^{-1}$, and specifically, the samples from Cosa exhibit a dual peak within this main Si-O bonding band (Figure 6) while it is less pronounced in the sample from Caesarea (Figure S5). The additional FTIR peaks at 890, 1390 and 1650 cm$^{-1}$ are typical of Si-O-Al, C-O and O-H bonding, respectively [37,38]. Subsampling within the Cosa cores did not result in a significant change in the spectra. To further explore the lime heterogeneity highlighted by the micro-CT imaging, we used micro X-ray fluorescence (micro-XRF) mapping of one of the relict lime clasts. It found the surface to be rich in calcium, sulfur, iron, potassium and aluminum (Figure 7 and S6).

Storage and transport properties were found to fall within two porosity-permeability trends separated into cores containing predominantly either mortar or aggregates. The concrete cores that are predominantly mortar (mortar-dominated) fall within porosity and permeability ranges of 45-55% and 1-5 10$^{-8}$ m$^2$, respectively (Figure 8). The Caesarea sample that falls outside of this trend was found to have debonding fractures that are responsible for the higher permeability. The porosity and permeability of the mortar-dominated samples match the values measured for the RMC analogous mortar samples made with low compression. Larger compression-derived compaction of the analogous mortar samples caused a linear decrease in the porosity - log permeability plot (Figure 8). The mortar-dominated samples exhibit lower permeabilities than the aggregate-dominated cores at similar porosities, reflecting differences in either the pore size distribution or connectivity, or both. It is worth noting that the aggregate-dominated samples from Baia and Portus Traiani contain volcanic tuff, the typical aggregate within many of the Roman marine concrete installations in Italy [4], resulting in porosity-permeability values that closely fit the trend of NYT [39]. Conversely, the aggregate-dominated sample from Caesarea has a much lower porosity and permeability, falling within the porosity-permeability trend traced by carbonate-rich grainstones with a grain size smaller than 20 μm [40]. The predominant aggregate used in RMC from Caesarea is the carbonate grainstone known locally in Palestine as *Kurkar* [4,41]. A reference sample is shown for concrete made from OPC (Ordinary Portland Cement) measured under similar conditions which falls between the two trends [42].

## 4. Discussion

High ductility and low permeability are two material properties used to achieve high durability in cementitious materials [21–23]. Fibers are commonly used to increase the strength and compressional-bulk resilience in cementitious and composite materials. The addition of macro-fibers and micro-fibers embedded in cement paste provides obstacles serving as crack arrestors, which contribute to retarding the propagation of cracks that lead to catastrophic failure [43]. The embedded nature of the sulfo-aluminate fibers found in RMC through SEM imaging, and the lack of evidence of late fiber formation, such as crack-filling, suggest a syngenetic formation with the matrix. The isolated micro-fracture development without coalescence, supports the interpretation of a ductile failure mode, which pertains more to soils and sediments than cemented materials. Furthermore, a pumice clast appears intact (Figure 2) despite being known to fail in volcanic tuffs under similar loading conditions [39], suggesting that the matrix is the load bearing phase. The overall rearrangement of the microstructure through pore collapse, grain rotation, and isolated micro-fracturing indicates that the intermolecular interactions within the matrix of RMC deform through a mechanism that absorbs mechanical energy. Studies in the literature show that matrices made of fibers also exhibit creep [44] and, in particular, the load-bearing fiber density is recognized to provide the scaffolding that is responsible for the transmission of forces from the macroscopic scale to the microscopic scale upon deformation [45] — in a network of entangled fibers, if one breaks others start resisting the deformation until they subsequently yield in a slow chain reaction. The embedded and intertwined fibers identified in this study may, too, contribute to the ductile deformation mode of RMC, increasing its durability.

While the addition of micro-fibrous structures, both natural and synthetic, is common within the concrete and composite industries to modify the strain absorption characteristics [46–48], the fiber-matrix interface is also recognized to be the key region that determines, to a great extent, the set of properties of composite materials. One of the mechanisms known to increase the bonding at the fiber-matrix interface is physical branching and chemical cross-linking leading to the entanglement of the fiber macromolecules with the matrix molecules, also known as polymerization of the matrix [49,50]. Therefore, any physical and chemical process that favors fiber growth and entanglement contributes to the reinforcement of the cementation process. The higher content of alkali in the *pozzolana* together with this content being inversely correlated with calcium content (Figure S3) prompted us to explore whether the matrix exhibits sign of an increased polymerization of alumino-silicates [51,52]. We used Fourier-transform infrared spectroscopy (FTIR) to compare the chemical bonding signatures of RMC with C-A-S-H, a geopolymer, and a C-A-S-H + geopolymer gel mixture [38,53,54]. Geopolymers describe a family of mineral binders that form through the chemical reaction of alumino-silicates with alkali polysilicates yielding polymeric Si-O-Al bonds [55]. Due to the high alkali and low calcium content of geopolymers, they are sometimes referred to as alkali-activated alumino silicate binders [56,57]. Dual Si-O bonding peaks are found to occur within mixtures of C-A-S-H and geopolymer gels where they are able to coexist [53]. The dual peaks in the Cosa samples align with both pure C-A-S-H and geopolymer spectrums, indicating that the silica and alumina are bound in both 2D and 3D structures within RMC, which tells us that polymerization has occurred. This is supported by the identification of both Q2(Al) and Q3(Al) peaks in solid Si Nuclear Magnetic Resonance (NMR) by Jackson et al. [11]. The formation of longer polymeric chains is mechanically relevant as it has been shown to control the strength and failure behavior. Gradin et al. [58] reports that at low levels of entanglement density (i.e., branched polymer), lower mechanical properties, particularly strength, are measured. As chains become longer, they become entangled, which leads to an increase in the length of the ductile plateau in the stress-strain curve due to chain sliding. As cross-linking increases there is also an increase in strength that accompanies plastic deformation beyond the yield point (i.e., strain hardening). The crossbred C-A-S-H-geopolymer matrix identified in this study further contributes to the ductile deformation mode of RMC by bonding the fibers and matrix, increasing its durability.

The porosity-permeability values of the mortar-dominated samples from Italy are consistent with and fall within the trend traced by RMC analogous mortar samples made under different confining pressures, which affects their densification (i.e. compaction). These samples have been produced without aggregates, confirming that (i) the matrix controls the flow while isolated aggregates being dispersed in the matrix of Roman concrete mortar do not significantly contribute to fluid flow and (ii) in the absence of cracks, the porosity-permeability relationship of the mortar-dominated samples is matrix-controlled. This would seem to contradict research in modern concretes, which shows that the addition of even small aggregates (~5 mm) results in a higher permeability than both the aggregates and cement paste, mostly due to cracking in the interfacial zone [59] that forms after setting. Nevertheless, the use of a compositionally-similar rubble aggregate such as Neapolitan Yellow Tuff (NYT) in RMC from Italy creates stronger bonding at the cement-aggregate interface due to the pozzolanic activity of NYT [4,60]. As a result, modern concrete measured under similar conditions falls above the porosity-permeability trend for the mortar-dominated RMC samples [42]. While the significantly lower porosity of modern concrete (~10%) contributes to it having a lower permeability than RMC, the superior aggregate bonding and polymerization of RMC provides a tight microstructure, thus resulting in a lower permeability than would be expected for concrete of a similar porosity. This limits the ability of water to permeate through the concrete, restricting the transport of potentially reactive ions, which promotes chemical weathering. This relationship does not hold for the mortar-dominated sample from Caesarea where the carbonate-rich grainstones that are used as a rubble aggregate, as opposed to the NYT used in the Italian RMC cores, has been reported to develop weaker bonds with the cement matrix resulting in interface failure causing debonding [41]. As a result of this weak bonding, Caesarea sample exhibit significant cracking, increasing the permeability. The restricted chemical weathering in the Italian RMC is highlighted by the relatively low measured chlorine (<7%) and sodium (<1%) concentrations despite being submerged in seawater for two millennia. Thus, we hypothesize that the sulfur-rich fibers finely interwoven in this crossbred matrix and cement-aggregate bonding contribute to a ductile deformation under stress as well as resistance to weathering and mechanical stress.

The presence of sulfur in the relict lime clasts, along with potassium and aluminum, is an unexpected finding and warrants further investigation as sulfur, aluminates, and alkalis have an impact in the chemical reactions with the pozzolan. Carbonate rocks by definition do not contain sulfur, which raises the question of the source material used in the production of lime and/or the water employed in its slaking process. The literature shows that limes made from pure calcium carbonate were little used in Roman construction because they were not of a hydraulic nature [61,62]. Dix [61] specifically highlights that, by containing sufficient impurities (more than 10%), hydraulic limes from marlstones allowed for slower slaking and setting underwater [62]. This would imply that rocks other than carbonates containing both calcium and alkalis were used (e.g., a local calc-alkaline volcanic rock such as trachyte-phonolites) [63], or alternatively, sulfur-rich and alkaline water from the volcanic manufacturing regions could be part of the making process [64,65]. This calls for a careful characterization of the relict lime clasts existing in the matrix of ancient Roman concrete and/or possibly retrieving the fluids used in the slaking process. The nature of these components may unveil a chemical process that the Ancients may have unwittingly exploited during the historical development of the mortar while possibly being inspired by the cementation of pozzolan-derived rocks from the calc-alkaline volcanic region of manufacture [16]. Understanding how materials behave under severe environments is a challenge as much for engineering as it is for geophysics. This cross-fertilization of knowledge can bring together solutions for understanding natural systems as well as devising Earth-inspired materials.

## Conclusions

Roman concrete has stood the test of time, showing remarkable chemical and physical resilience. We investigated how the microscopic physical and chemical heterogeneity influences the macroscopic failure behavior by coupling the mechanical stressing of Roman marine concrete (RMC) with multi-scale and multi-dimensional imaging, including chemical mapping and time-lapse 3-D. We found that RMC creeps and exhibits a ductile fracture behavior and that the formation of a matrix-aggregate interfacial zone is limited by the use of compositionally similar aggregates, lowering permeability. Structural and chemical imaging showed a microstructure of an intertwined network of calcium-sulfo-aluminate fibers embedded in a crossbred matrix made of calcium-aluminum-silicate-hydrates and a geopolymer. Both the fibers and relict lime are particularly rich in sulfur and alkalis, likely reflecting the ultra-alkaline volcanic nature of the primary source materials available in the manufacturing region. The richness in alkalis is particularly important as it enhances the cohesion at the fiber-matrix interface through polymerization. Furthermore, the sulfur- and alkali-richness of the lime challenges the current wisdom of a lime-producing rock of pure carbonate origin, which warrants further investigation as it opens up the possibility of cross-fertilization between the fields of engineering and geophysics.

Table 1: Composition of natural (top) and industrial (bottom) pozzolans used in historical and modern cement production. It is worth noting the large amount of alkalis in natural pozzolans from calc-alkaline to ultra-alkaline volcanic regions above subduction zones. Pozzolana and Santorini Earth values were measured in this study.

| Material | $SiO_2$ | $Al_2O_3$ | $Na_2O$ | $K_2O$ | $SO_3$ | $MgO$ | $CaO$ | $FeO/Fe_2O_3$ |
|---|---|---|---|---|---|---|---|---|
| Pozzolana | 52.1 | 14.0 | 1.7 | 6.9 | 0.1 | 1.0 | 2.4 | 3.8 |
| Santorini Earth | 67.9 | 11.7 | 2.5 | 2.9 | 0 | 0.6 | 2.2 | 3.2 |
| Shirasu [66] | 69.3 | 14.6 | 3.0 | 2.4 | | | 2.6 | 1.0 |
| Iwo Jima Ash [67] | 60.4 | 16.7 | 6.0 | 4.0 | | 1.5 | 3.4 | 6.5 |
| Class F Fly Ash [68] | 62 | 18.9 | 2.4 | 1.1 | | 2.0 | 6.0 | 4.9 |
| Ground Slag [68] | 33 | 13.4 | 0.2 | 0.4 | | 6.0 | 41.8 | 0.2 |

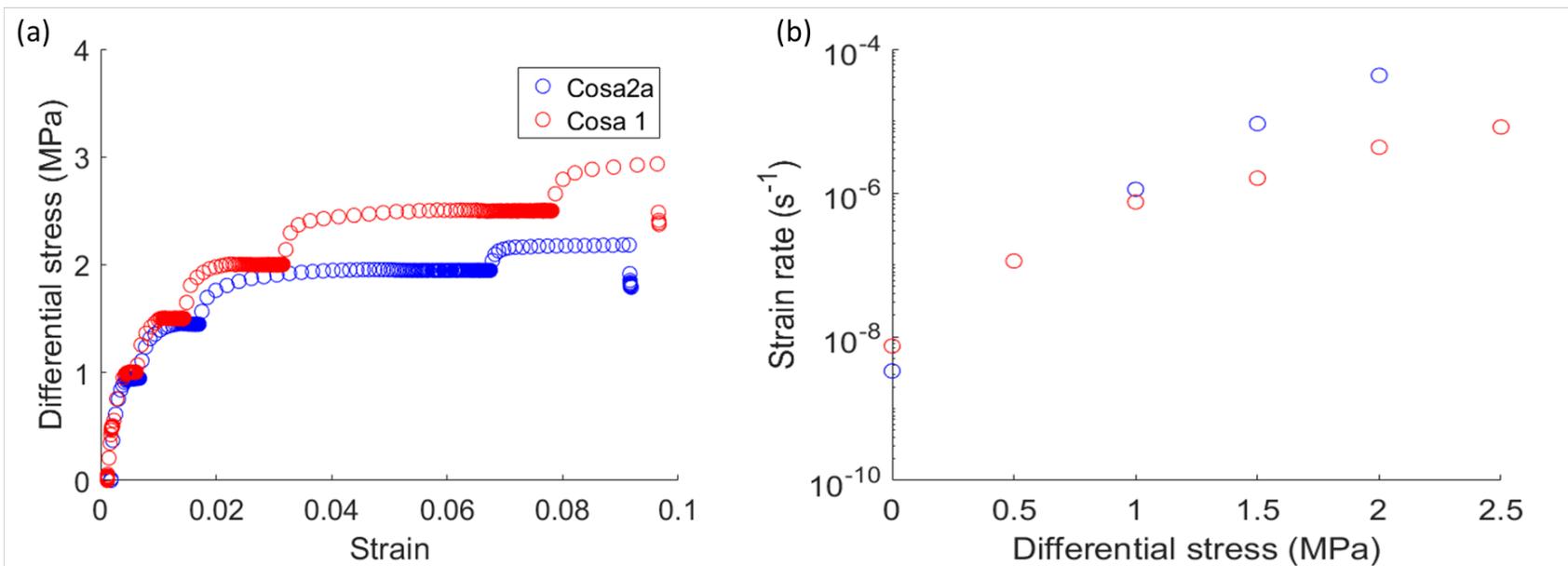

Figure 1: (a) Differential stress as a function of strain during triaxial stress stepping measurements (radial stress of 3 MPa). Time between data points is 7 seconds. (b) Strain rate as a function of differential stress during triaxial loading. Radial stress is held constant at 3 MPa. The red and blue dots represent samples Cosa 1 and Cosa 2a, respectively, matching (a).

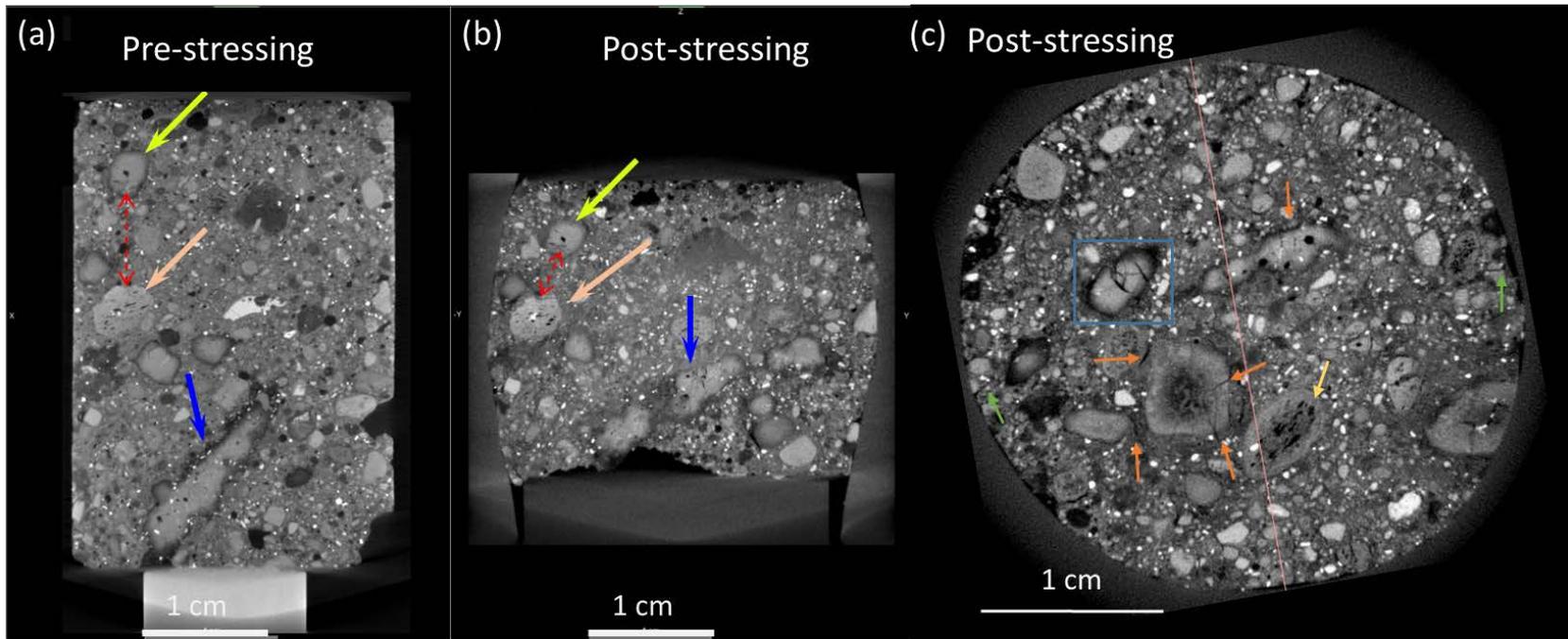

*Figure 2: (a-b) Vertical cross sections of the pre- and post-stressing micro-CT scans of Cosa2. Both images have the same scale and contrast/brightness. The blue and yellow arrows indicate relict lime clasts and the pink arrow indicates a pumice aggregate that can be correlated between the two images. Rotation and fracturing are seen in the relict lime while the pumice stays intact. The red arrow highlights the relative motion of the features. Horizontal cross-sections can be found in the supporting materials and highlight the micro-fractures that are seen within the matrix. (c) Horizontal microCT cross-section through a sample from after triaxial stressing. The image shows isolated cracking within the relict lime clasts (blue box), at the aggregate-cement interface (orange arrows) and radially around the sample (green arrows) which are not present in the pre-stressed image. A vescicular pumice clast (yellow arrow) appears intact despite being known to crash under loading conditions [39]. A pre-stressing cross-section is shown in Figure S8.*

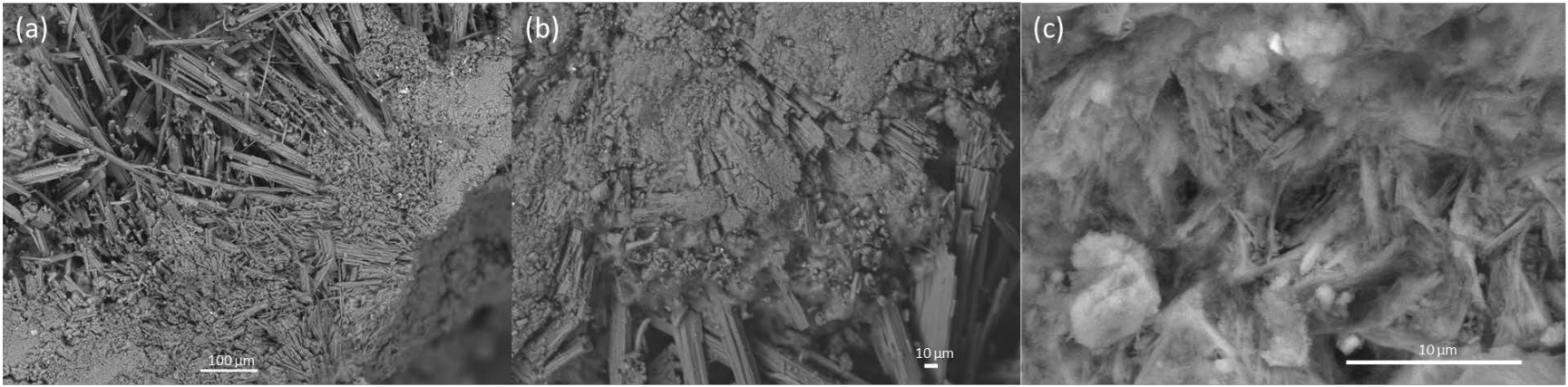

*Figure 3: (a, b) SEM-BSE (backscatter electron) images of pervasive, embedded fibers that sprout from the matrix and bridge the pore space in samples from Cosa. (c) SEM-BSE image of fibers intertwined within the matrix from Baia.*

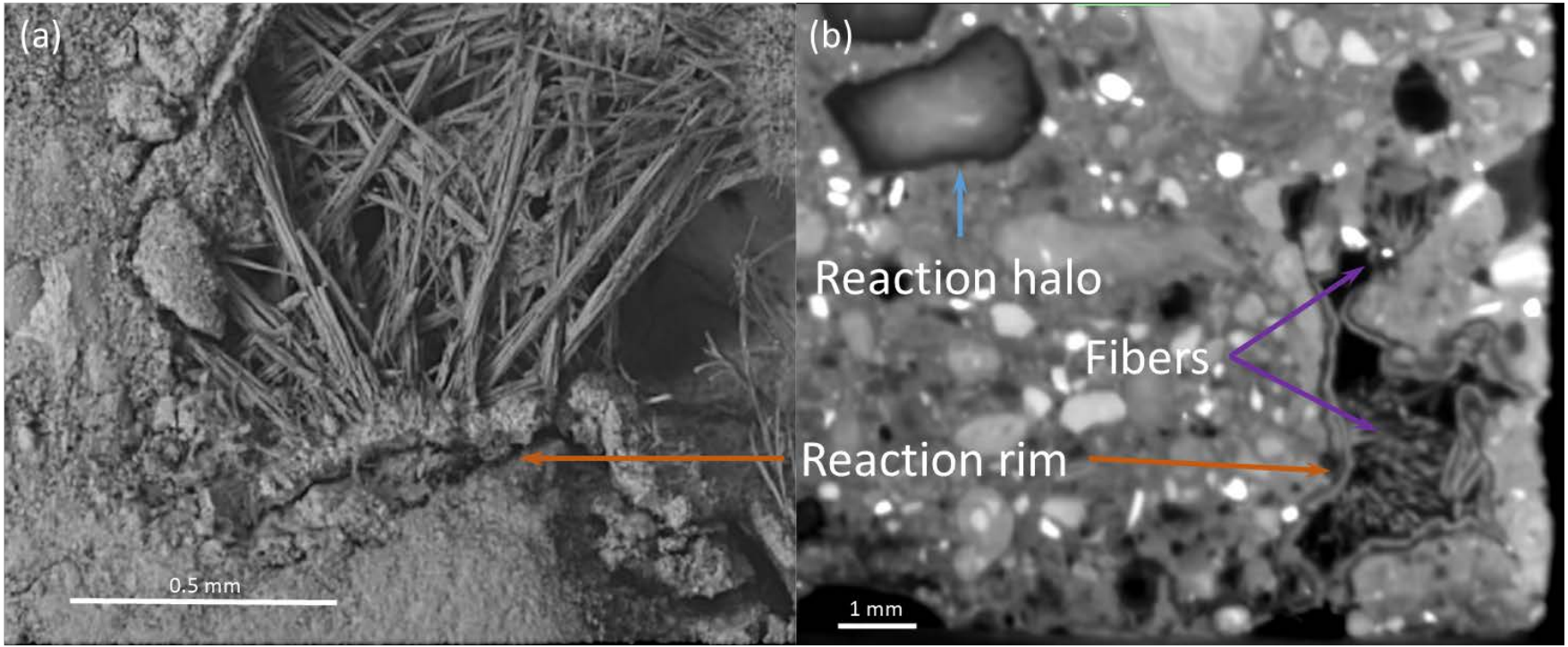

*Figure 4: (a) SEM-BSE image of a fiber filled relict lime clast in a sample from Cosa with the surrounding reaction rim highlighted. (b) Micro-CT slice through the same sample that shows the two types of relict lime clasts: fiber filled pores (purple arrows) with reaction rims (orange arrows) and dense, filled pores with reaction halos (blue arrow).*

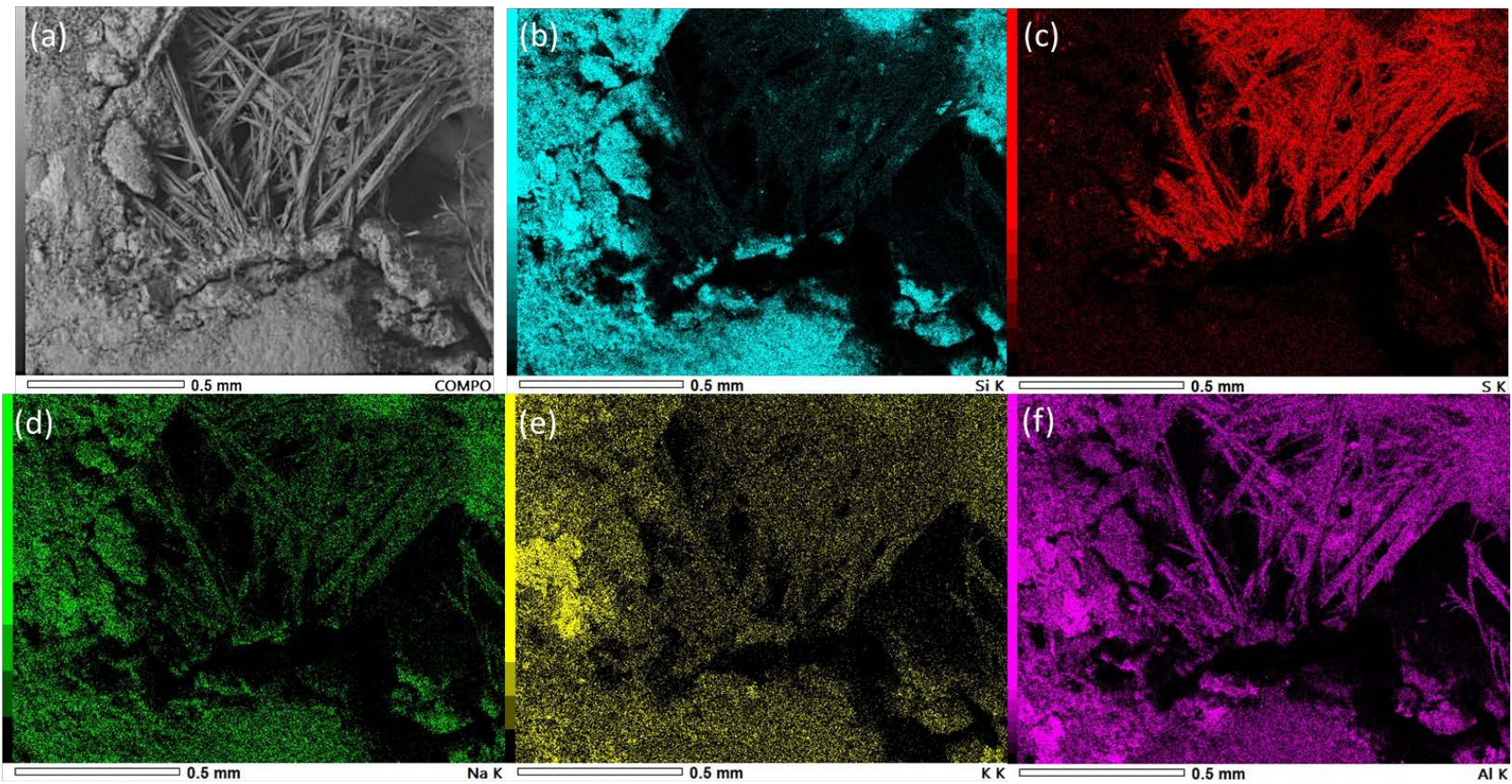

Figure 5: (a) SEM-BSE image of a fiber filled relict lime clast in a sample from Cosa. SEM-EDS (energy dispersive) imaging shows that the matrix is rich in Si (b) whereas the fibers are rich in S (c). Both the fibers and matrix are rich in Na (d), K (e), Al (f) and calcium (Figure S3).

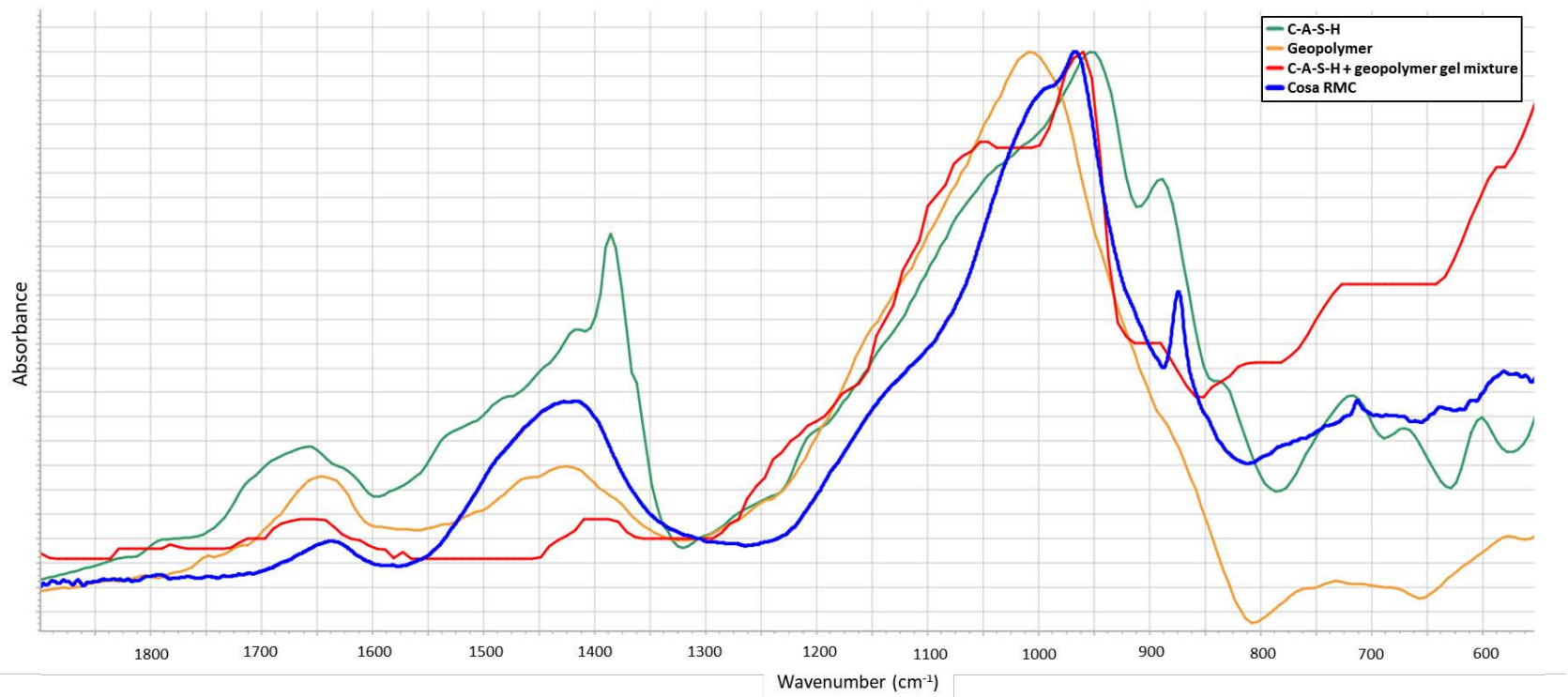

*Figure 6: FTIR spectra for a sample from Cosa (blue), a C-A-S-H and geopolymer (N-A-S-H) gel mixture (red) [53], C-A-S-H (Ca/Si 1.0 and Al/Si 0.2) (green) [38], geopolymer (alkali activated pozzolan) (orange) [54]. The Roman concrete spectrum falls between the three waveforms and shares similar peaks. The absorbance has been normalized for comparison.*

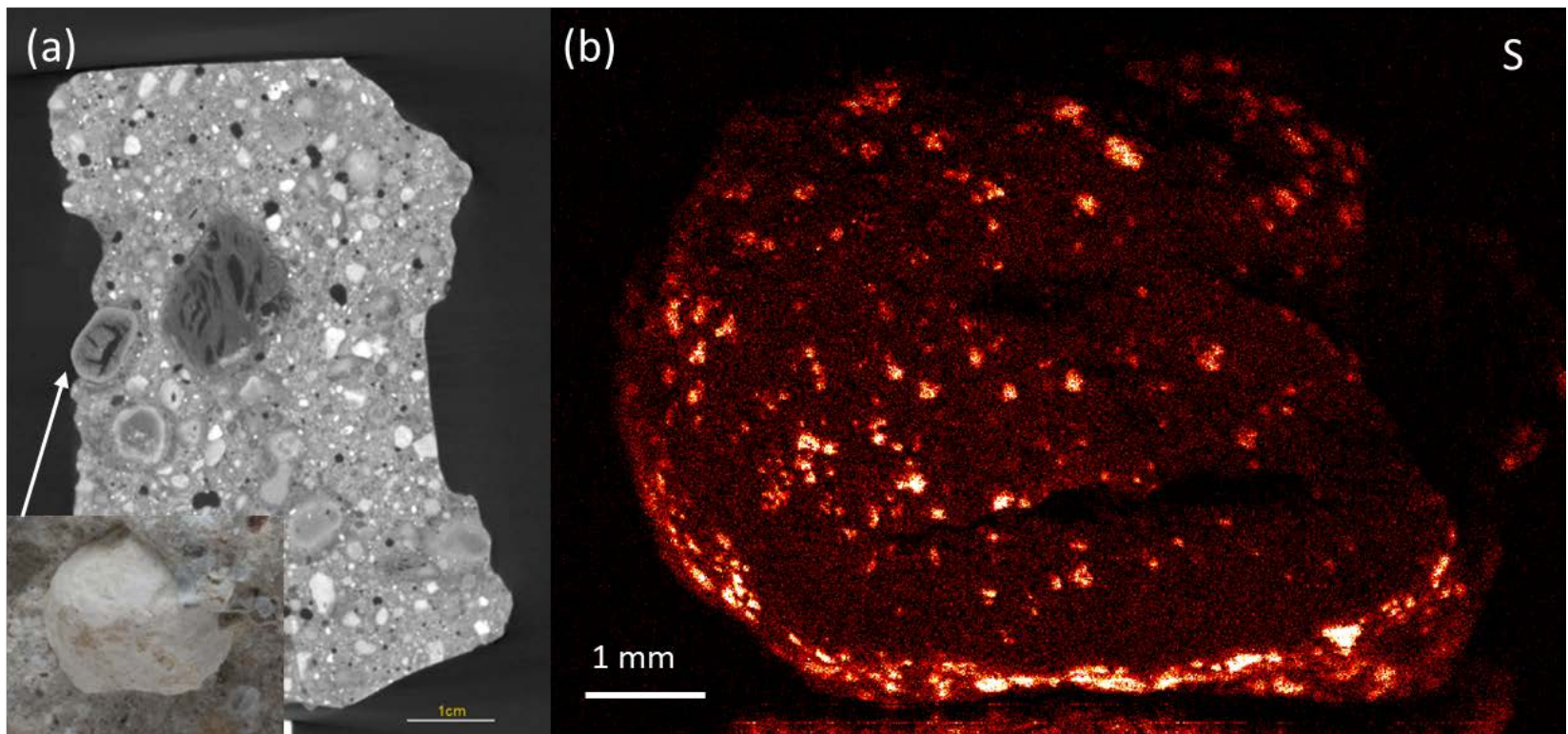

*Figure 7: CT reconstruction showing the relict lime clast of interest (optical image inset). (b) Surface micro-XRF of the relict lime clast showing the intensity of sulfur concentration. The energy spectra for two regions of interest are shown in Figure S6.*

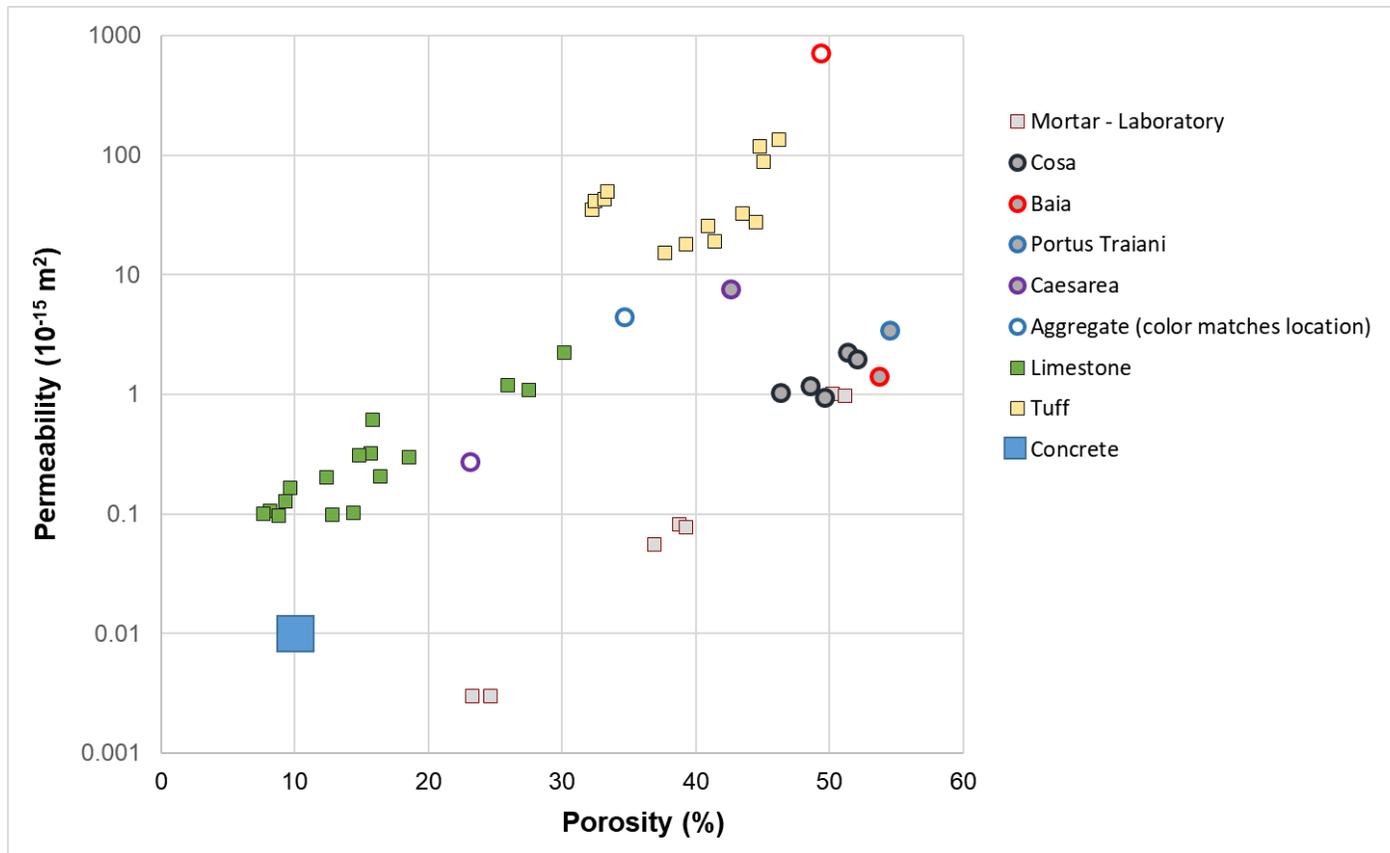

Figure 8: Measured helium porosity and permeability data. The Roman concrete cores are separated into mortar- and aggregate-dominated groups, which show two distinct trends. The mortar-dominated samples from Cosa (black), Baia (red) and Portus Traiani (blue) correlate with laboratory lime-pozzolan mortar samples (red outline squares), with increasing compaction displaying a trend parallel to the aggregate samples. The aggregate-dominated samples from Baia and Portus Traiani (red and blue open circles) contain volcanic tuff and closely fit the trend of Neapolitan Yellow Tuff [39] (yellow squares). The aggregate-dominated sample from Caesarea (purple open circle) fall within the porosity-permeability trend traced by limestone (green squares), being one of the typical aggregates used in concrete from this location [4,40,41]. The limestone samples are a subsample of Lucia [40] with grain sizes smaller than 20 µm. The Caesarea sample (purple) was found to have significant debonding fractures which are responsible for its higher permeability. A reference range is shown for comparably measured Portland cement concrete (blue square) [42].

**Acknowledgements**

The authors would like to acknowledge the guidance and input of Prof. Luigi Biolzi and Dr. Kyle Douglas. We would also like to thank for their technical support Anthony C. Clark, Arturas Vailionis, Dale Burns, SIGRAY and Zeiss.

This work was supported by the NSF CAREER Award (EAR-1451345 to T.V.) and by the startup fund to T.V. of the School of Earth, Energy, and Environmental Sciences at Stanford University. PM acknowledges the financial support given by the Roy W. Carlson Chair.

Part of this work was performed at the Stanford Nano Shared Facilities (SNSF) using the Zeiss Xradia 520 Versa, acquired with support from NSF as part of CMMI-1532224. SNSF is supported by the NSF as part of the National Nanotechnology Coordinated Infrastructure under award ECCS-1542452.

**Additional information**

The authors declare no competing interests, neither financial nor non-financial.

# Supplementary Information for

**The Importance of the Ultra-alkaline Volcanic Nature of the Raw Materials to the Ductility of Roman Marine Concrete**

Jackson MacFarlane[1], Tiziana Vanorio[1] and Paulo J.M. Monteiro[2]

[1]Stanford Rock Physics Laboratory, Stanford University

[2]Department of Civil Engineering, UC Berkeley

**Correspondence to:** Jackson MacFarlane, email: jmacfarl@stanford.edu

**This PDF file includes:**

Table S1 to S2
Figs. S1 to S8

*Table S1: RMC analogous mortar samples preparation, including mass percentages of ingredients, time in steam chamber and compression applied. NYT = Neapolitan Yellow Tuff.*

| Sample | Pozzolana (%) | Lime (%) | Water (%) | NYT (%) | Cure time | Compression |
|---|---|---|---|---|---|---|
| **D6** | 48 | 24 | 29 | 0 | 28 | 5 lb mass for 72 hours |
| **D8** | 52 | 26 | 22 | 0 | 28 | 5 lb mass for 72 hours |
| **D5** | 48 | 24 | 29 | 0 | 28 | 2 MPa for 24 hours |
| **D7** | 52 | 26 | 22 | 0 | 28 | 2 MPa for 24 hours |
| **D11** | 52 | 26 | 22 | 0 | 14 | 2 MPa for 3 hours, 5 lb mass for 21 hours |
| **D12** | 52 | 26 | 22 | 0 | 14 | 2 MPa for 3 hours, 5 lb mass for 21 hours |
| **D13** | 52 | 26 | 22 | 0 | 14 | 2 MPa for 3 hours, 5 lb mass for 21 hours |

*Table S2: EDS point analysis of points 1, 2 and 3 as indicated in Figure S7. While both are rich in Ca, Al and O the Si and S contents are significantly different.*

|    | Point 1 (mass %) | Point 2 (mass %) | Point 3 (mass %) |
|----|------------------|------------------|------------------|
| O  | 14.43            | 50.35            | 46.23            |
| Na | 0.22             | 0.44             | 0.78             |
| Mg | 0.60             | 0.37             | 0.61             |
| Al | 7.69             | 2.32             | 3.33             |
| Si | 1.14             | 9.25             | 17.69            |
| S  | 8.22             | 0.31             | 0                |
| Cl | 6.80             | 1.28             | 2.05             |
| Ca | 60.9             | 35.69            | 27.70            |
| K  | 0                | 0                | 1.60             |

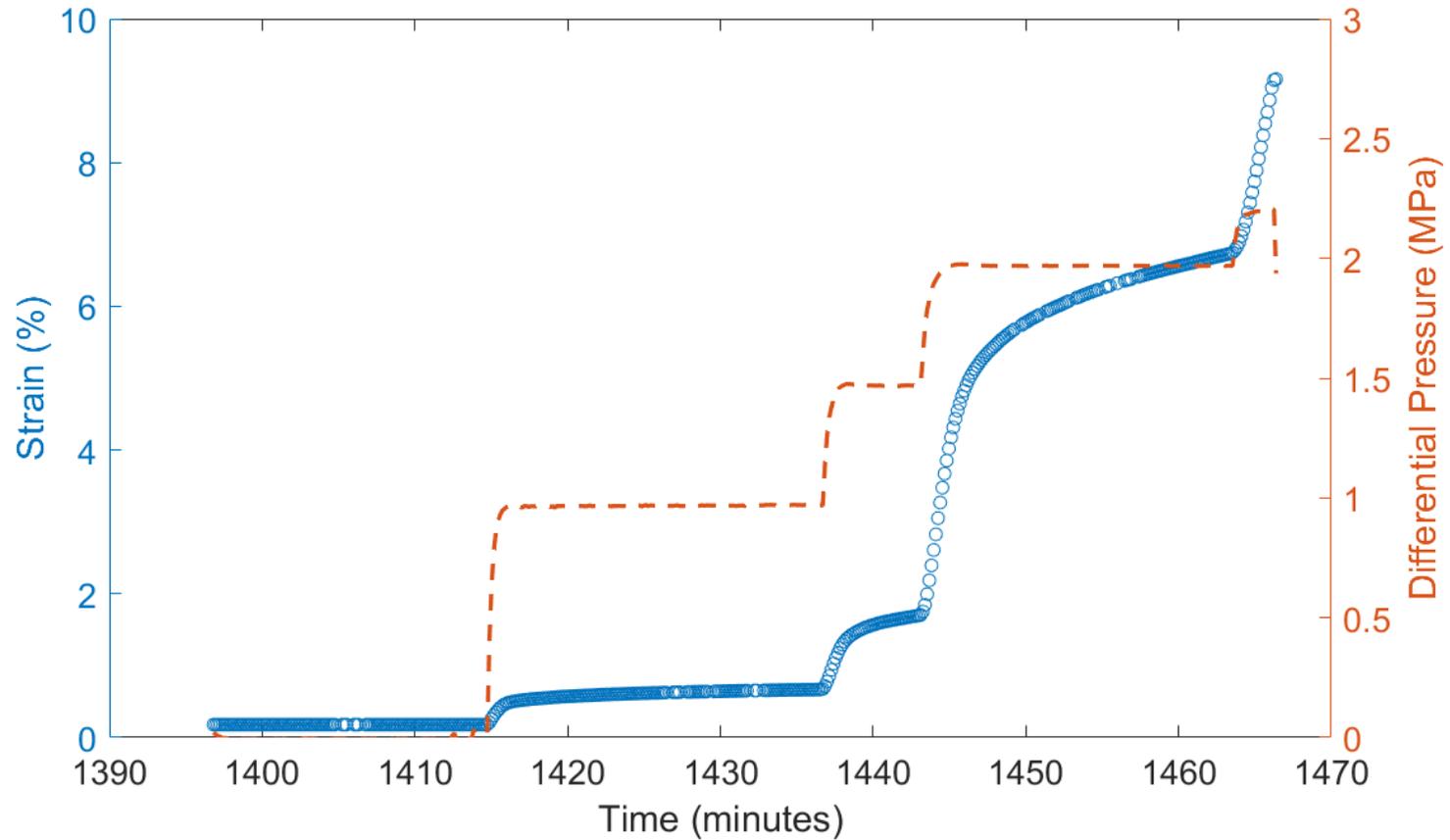

*Figure S1: Axial strain (blue) and axial deviatoric (differential) pressure (orange) as a function of time during triaxial loading of sample Cosa2a. Creep can be seen at the constant pressure intervals as the strain continues to increase despite constant loading. The increasing strain rate is visible in the steeper slope as the differential pressure increases.*

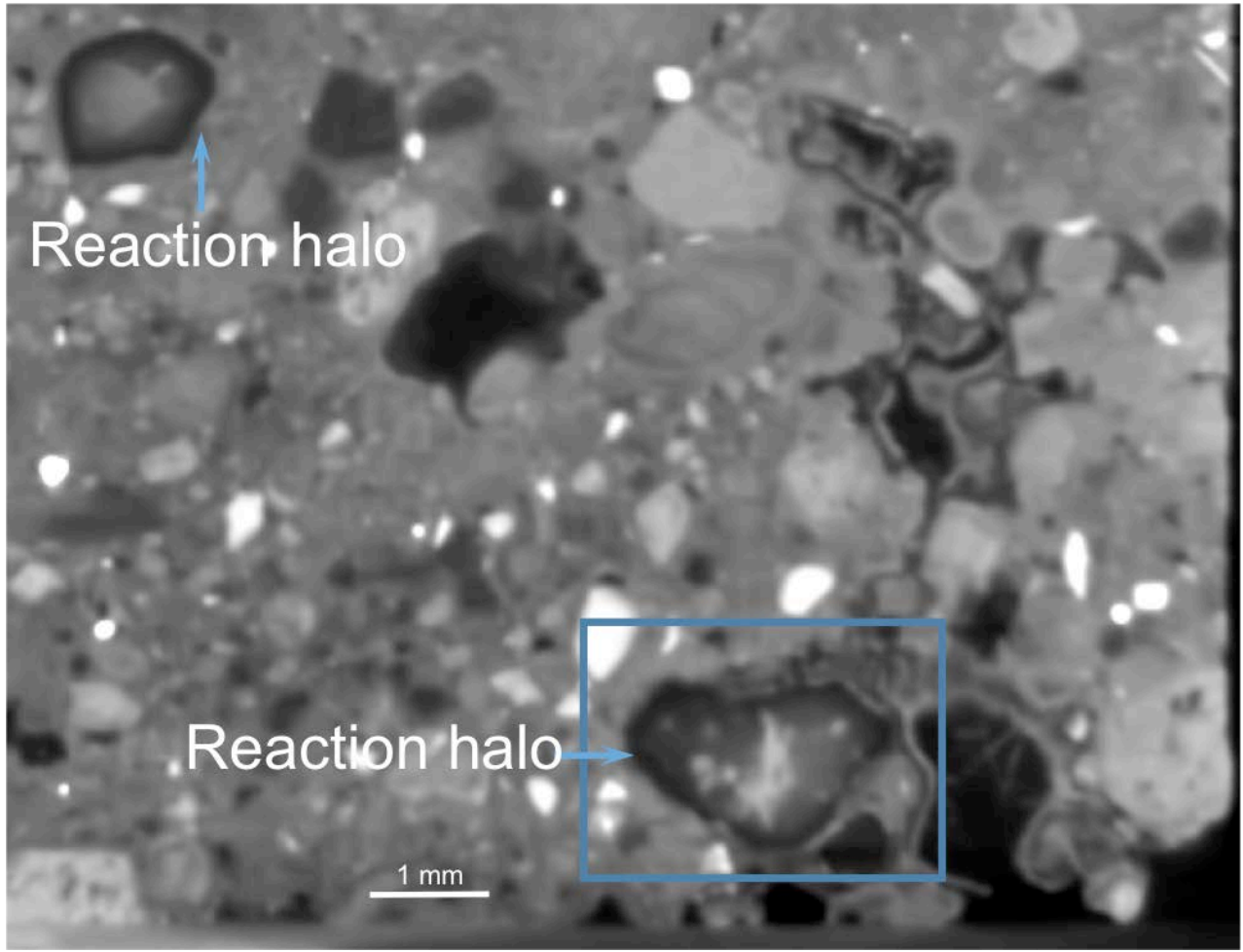

*Figure S2: MicroCT slice through a sample from Cosa, which highlights the internal heterogeneity within a relict lime clast (blue box) that has reaction halo around it.*

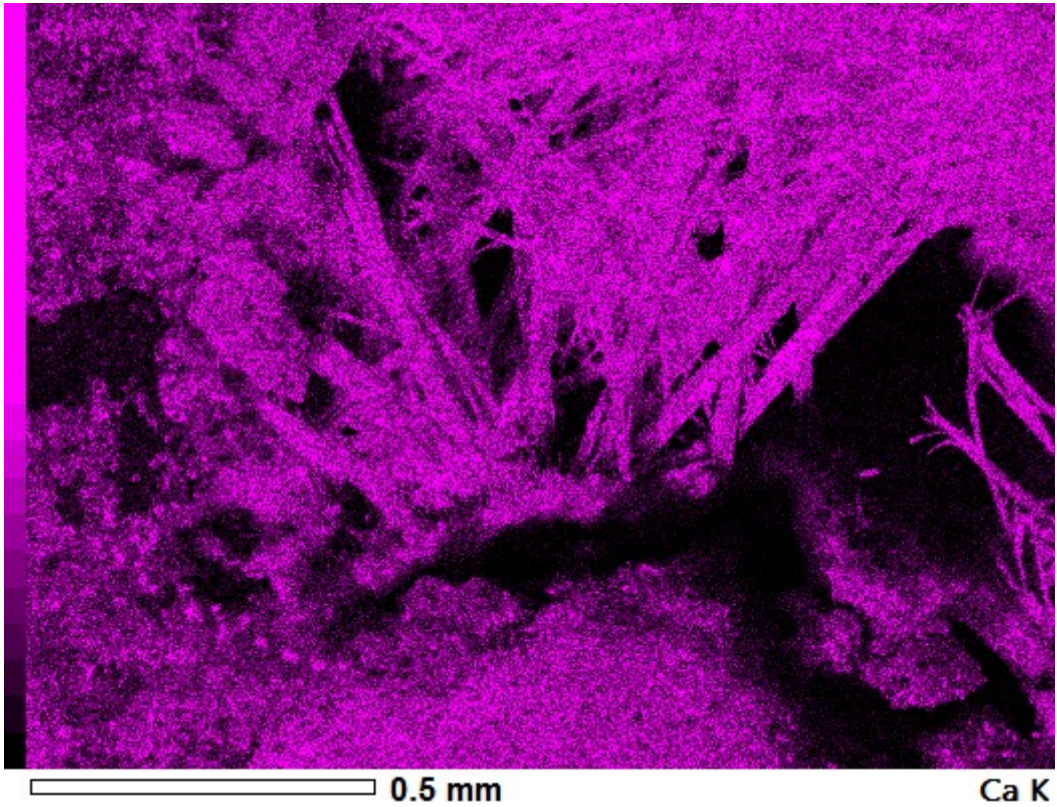

*Figure S3: SEM-EDS (energy dispersive) imaging shows that the both the matrix and fibers are rich in calcium.*

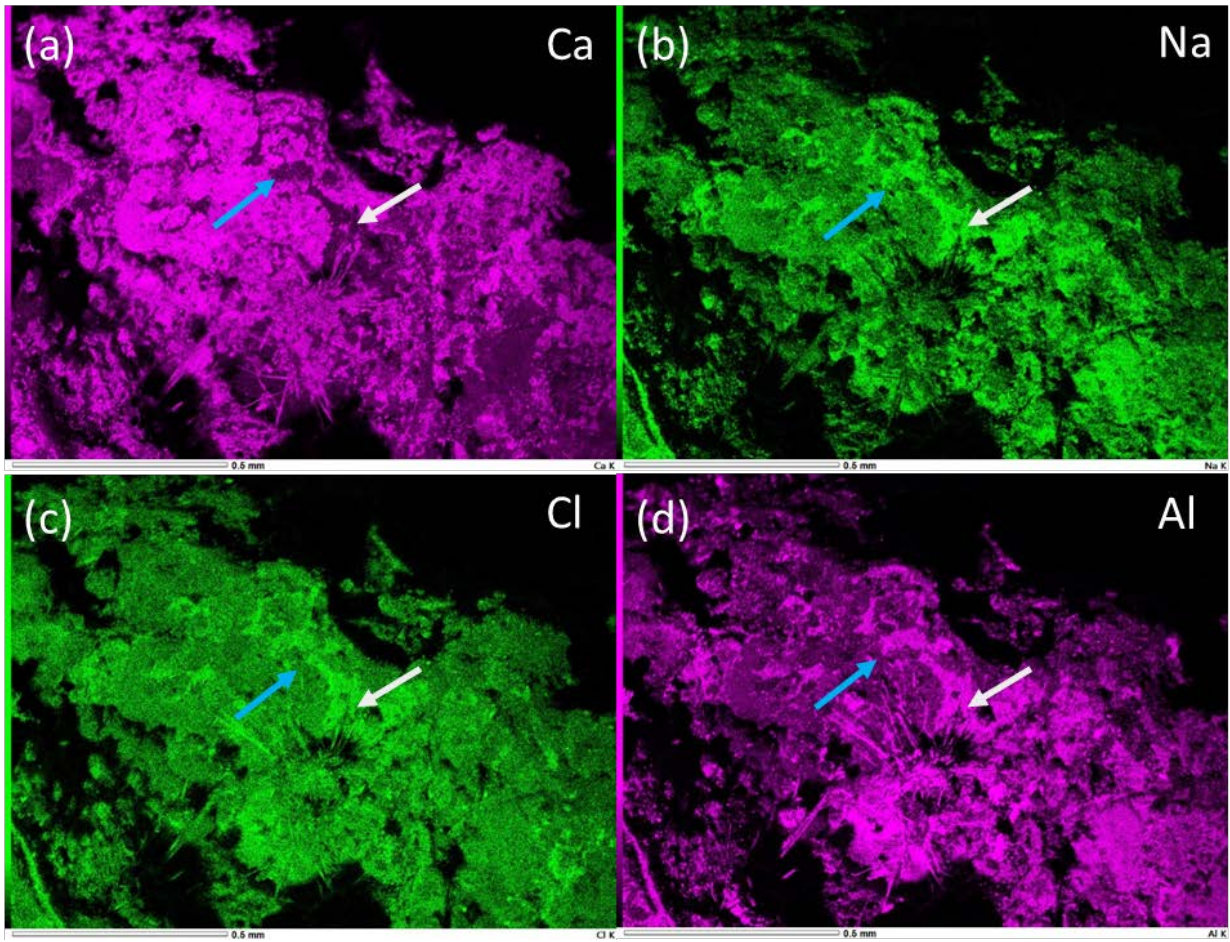

*Figure S4: SEM-EDS (energy dispersive) imaging shows a negative correlation between calcium (a) and the alkalis sodium, chlorine and aluminum (b-d) in the matrix of a sample from Baia. The white arrow indicates a low calcium, high alkali region while the blue arrow indicates a high calcium, low alkali region.*

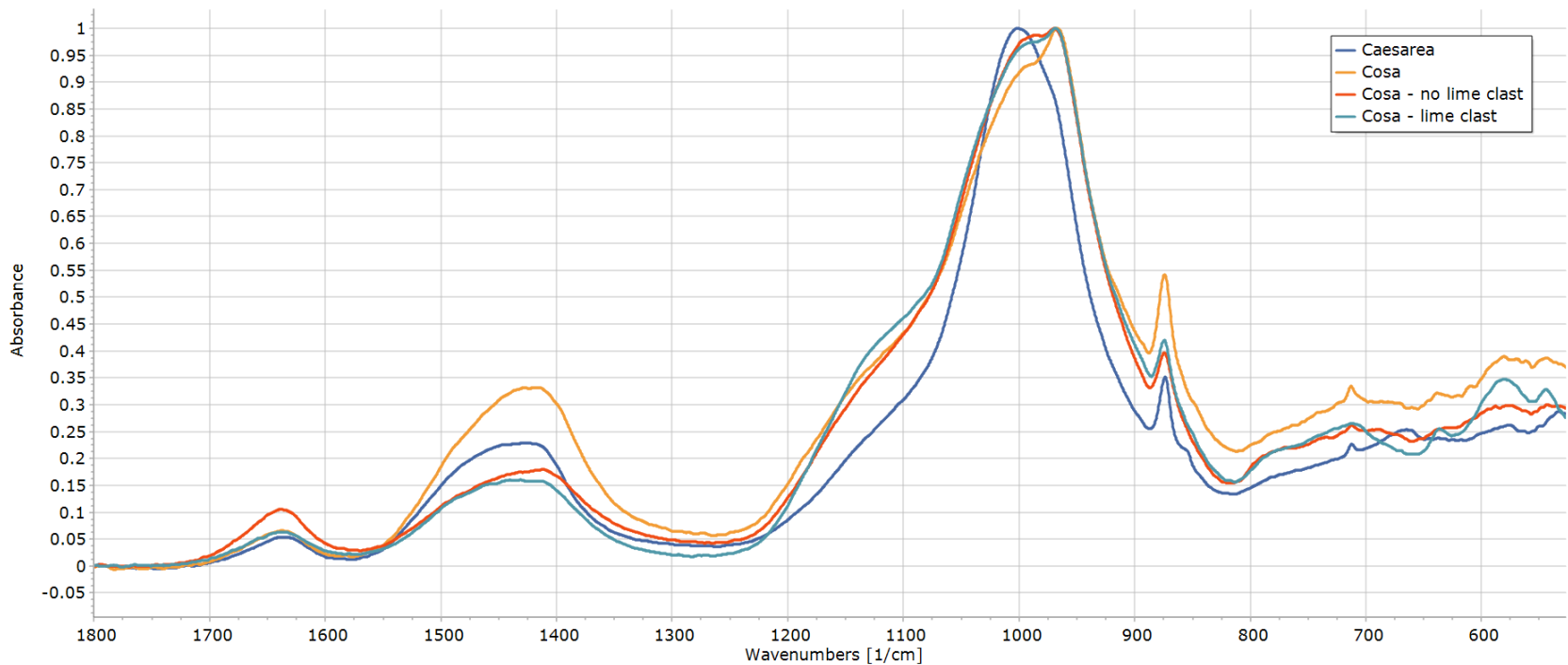

Figure S5: FTIR spectra for a samples from Cosa and Caesarea. All of the cores from Cosa from Cosa have a dual-peak between 1050 and 950 cm$^{-1}$. Sub-sampling within the same Cosa core (with and without a large lime clast, as specified in the Methods section) did not result in a significant change in spectra. Absorbance has been normalized by the highest peak for comparison.

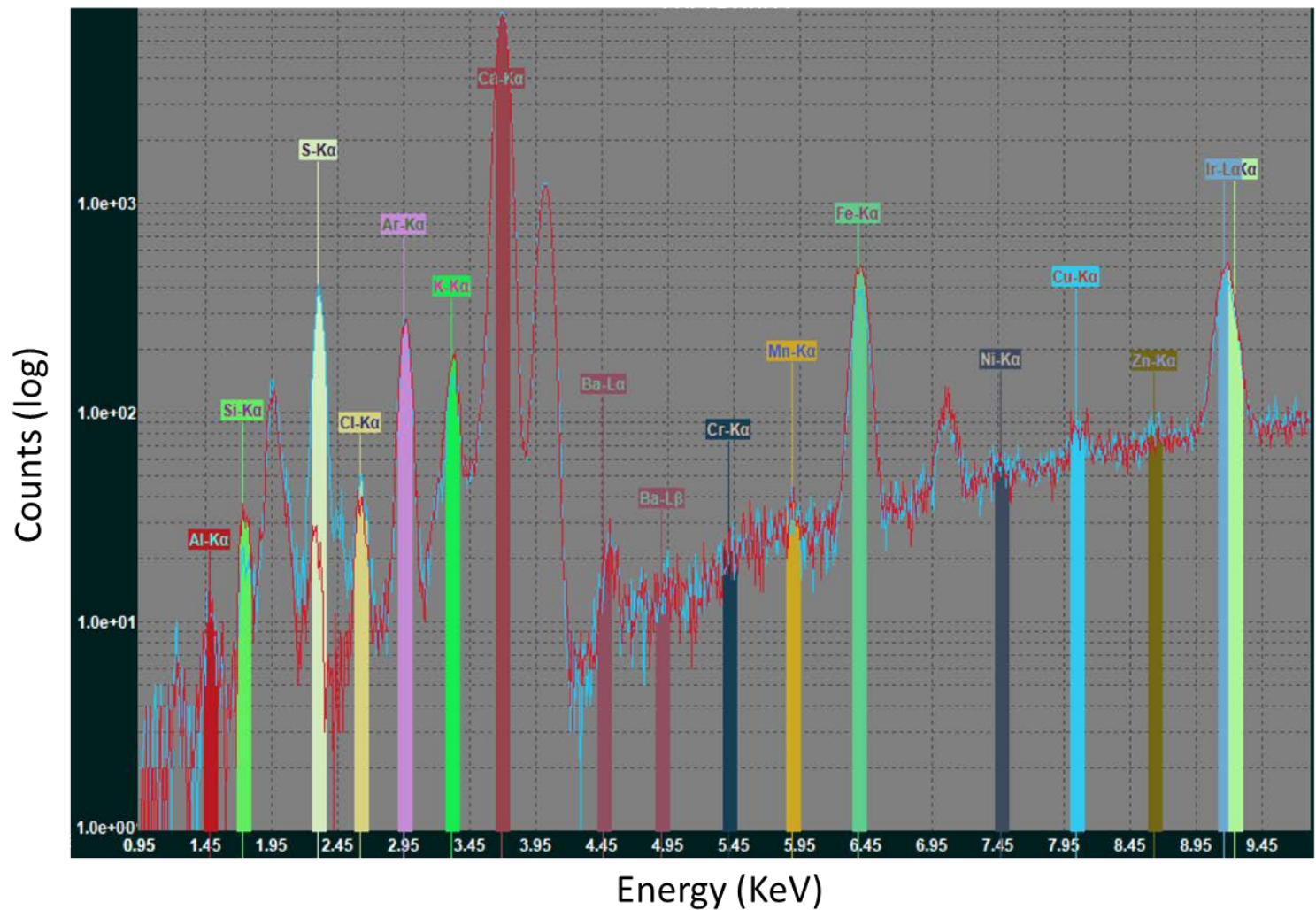

*Figure S6: micro-XRF energy counts for two region of interests that showed high (blue) and low (red) intensity on the sulfur map in Figure 7. The elemental correlation for each energy level is indicated by the colors bars. Both regions are rich in calcium, iron and potassium with only the sulfur content varying between the two locations.*

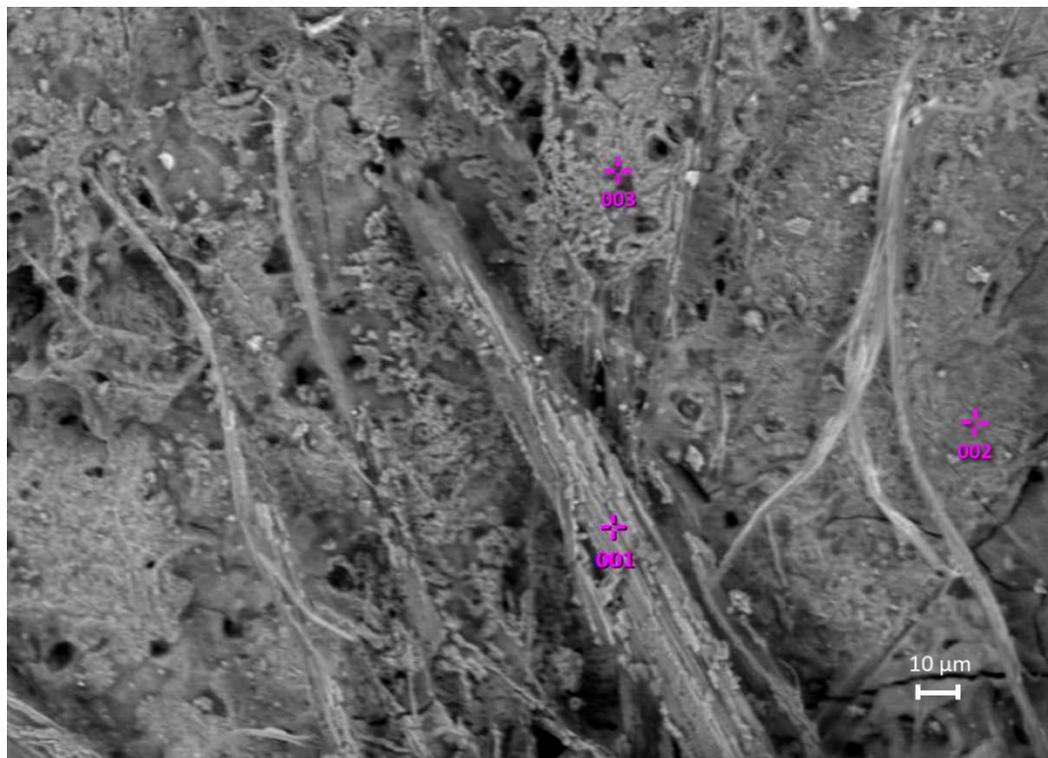
Figure S7: SEM image of embedded fibers within a Baia core sample. The numbered points show the locations of EDS analyses with the results given in Table S2.

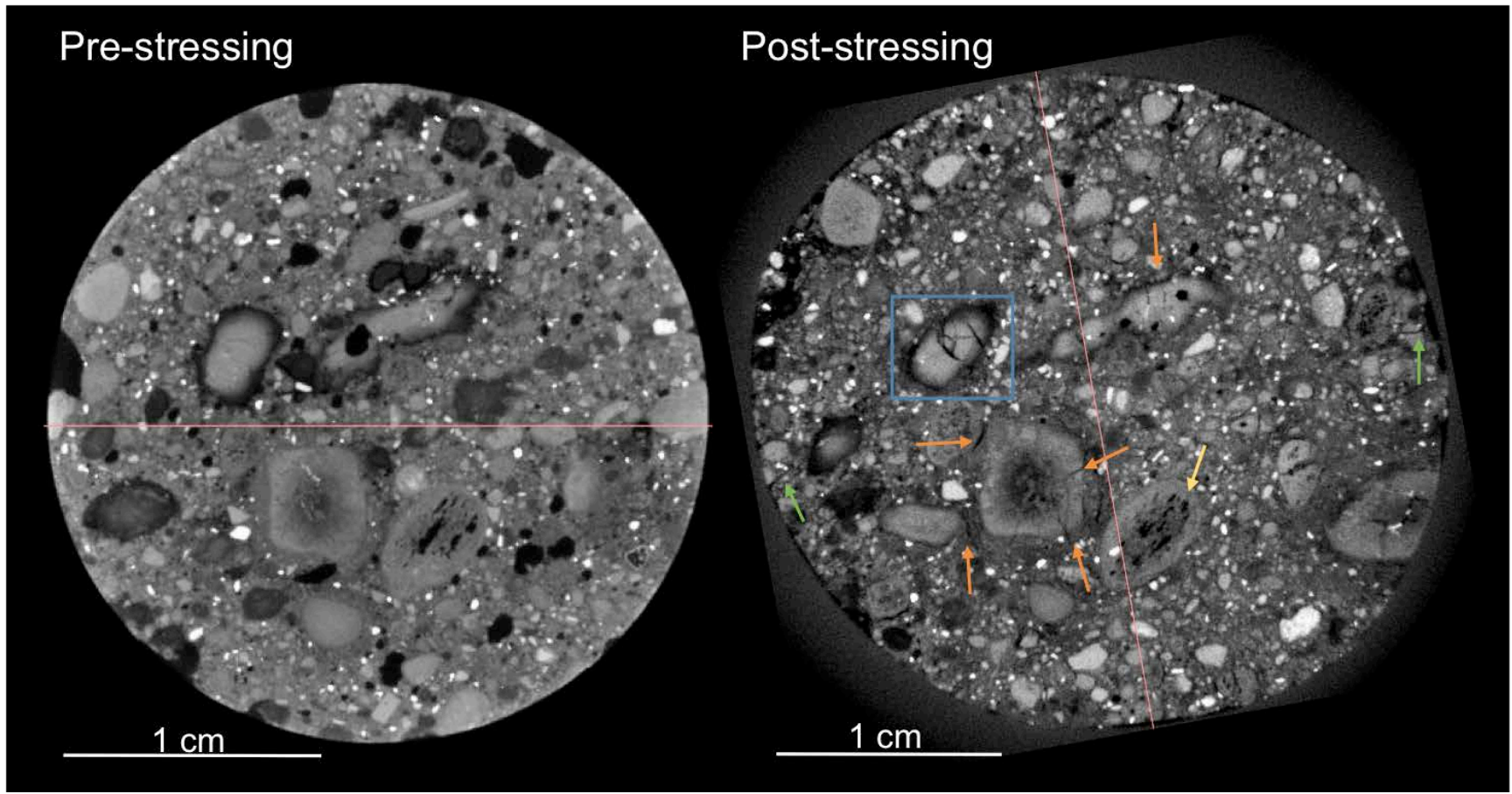

*Figure S8: Horizontal microCT cross-sections through a sample from Cosa before and after triaxial stressing. The post-stressing sample shows isolated cracking within the relict lime clasts (blue box), at the aggregate-cement interface (orange arrows) and radially around the sample (green arrows) which are not present in the pre-stressed image. A vescicular pumice clast (yellow arrow) appears intact despite being known to crash under loading conditions* [39].